\documentclass[reprint,superscriptaddress,showpacs,aps,pre,floatfix,longbibliography]{revtex4-1}

\usepackage{enumitem}
\usepackage{graphicx}

\usepackage{dcolumn}
\usepackage{color}
\usepackage{bm}
\usepackage[utf8]{inputenc}
\usepackage{xcolor}

\definecolor{darkred}{rgb}{0.6, 0, 0}
\usepackage[colorlinks=true, linkcolor=darkred, citecolor=darkred, urlcolor=darkred, filecolor=darkred]{hyperref}

\begin{document}

\title{The Physics of Life: Exploring Information as a Distinctive Feature of Living Systems}

\author{Stuart Bartlett}
\affiliation{Division of Geological and Planetary Sciences, California Institute of Technology, Pasadena, CA, USA}

\author{Andrew W. Eckford}
\affiliation{Department of Electrical Engineering and Computer Science, York University, Toronto, ON, Canada}

\author{Matthew Egbert}
\affiliation{School of Computer Science, University of Auckland, Auckland, New Zealand}

\author{Manasvi Lingam}
\affiliation{Department of Aerospace, Physics and Space Sciences, Florida Institute of Technology, Melbourne, FL, USA\looseness=-1}
\affiliation{Department of Chemistry and Chemical Engineering, Florida Institute of Technology, Melbourne, FL, USA\looseness=-1}
\affiliation{Department of Mathematics and Systems Engineering, Florida Institute of Technology, Melbourne, FL, USA\looseness=-1}
\affiliation{Department of Physics, The University of Texas at Austin, Austin, TX, USA}

\author{Artemy Kolchinsky}
  \email[Correspondence email address: ]{artemyk@gmail.com}
\affiliation{ICREA-Complex Systems Lab, Universitat Pompeu Fabra, Barcelona, Spain}
\affiliation{Universal Biology Institute, The University of Tokyo, 7-3-1 Hongo, Bunkyo-ku, Tokyo 113-0033, Japan\looseness=-1}

\author{Adam Frank}
  \email[Correspondence email address: ]{afrank@pas.rochester.edu}
\affiliation{Department of Physics and Astronomy, University of Rochester, Rochester, NY, USA}

\author{Gourab Ghoshal}
  \email[Correspondence email address: ]{gghoshal@pas.rochester.edu}
\affiliation{Department of Physics and Astronomy, University of Rochester, Rochester, NY, USA}
\affiliation{Department of Computer Science, University of Rochester, Rochester, NY, USA}

\begin{abstract}
This paper explores the idea that information is an essential and distinctive feature of living systems. Unlike non-living systems, living systems actively acquire, process, and use information about their environments to respond to changing conditions, sustain themselves, and achieve other intrinsic goals. We discuss relevant theoretical frameworks such as ``semantic information'' and ``fitness value of information''. We also highlight the broader implications of our perspective for fields such as origins-of-life research and astrobiology. 
In particular, we touch on the transition to information-driven systems as a key step in abiogenesis, informational constraints as determinants of planetary habitability, and informational biosignatures for detecting life beyond Earth. We briefly discuss experimental platforms 
which offer opportunities to investigate these theoretical concepts in controlled environments. By integrating theoretical and experimental approaches, this perspective advances our understanding of life's informational dynamics and its universal principles across diverse scientific domains.\end{abstract}

\maketitle

\section{Introduction}

Living systems actively sustain and renew themselves despite the natural tendency toward decay, a process sometimes termed \textit{autopoiesis} in the literature~\cite{maturana_autopoiesis_1980,varela_principles_1979a,frank_blind_2024a}. Recent research has investigated this concept within broader frameworks of cognition and adaptive behavior~\cite{beer_cognitive_2014,barandiaran_normestablishing_2013,beer_theoretical_2023,PL15,BL16,JAS21}. Central to this research is the idea that living systems are \emph{agents} that possess intrinsic goals~\cite{barandiaran2009defining}, such as \emph{viability} (maintenance of the living state), growth, and replication. In fact, the presence of goals that are intrinsic, rather than externally assigned, distinguish organisms from most non-living systems considered in the natural sciences~\cite{kauffman2019world,ball2023life}. 

One important way that organisms achieve intrinsic goals is by changing their behavior in response to different environments~\cite{egbert2023behaviour,kolchinsky2018semantic}. In this way, organisms acquire, process, and use information about environmental states for functional purposes. This kind of information usage has been variously termed ``functional''~\cite{collier2008information}, ``meaningful''~\cite{nehaniv2002meaningful}, and ``semantic''~\cite{kolchinsky2018semantic} information in the literature (with subtle differences). 

In this piece, we argue that the use of semantic information is one of the most distinctive and important features of living and proto-living systems~\cite{Schlosser_1998,SS00,Mossio_2009,godfrey-smithBiologicalInformation2007,gatenbyInformationTheoryLiving2007,walker2013algorithmic}.  In other words, although many non-living systems are usefully described by information-theoretic measures, organisms are distinguished by actively using information to sustain viability. We term this the \emph{informational perspective}.

The informational perspective suggests specific research directions, e.g., quantifying goal-directed information processing, understanding its role in the origin of life, and detecting it in astrobiological settings. The perspective is complementary to existing research in biology and biophysics that uses information theory to study information flows in modern organisms~\cite{FNG13,TB16,donaldson2010fitness,frank2012natural,schneiderEvolutionBiologicalInformation2000,gatenbyInformationTheoryLiving2007,ulanowiczInformationTheoryEcology2001,RCL12,adamiInformationTheoryMolecular2004,CA24,bialek2012biophysics}. It is also complementary to other perspectives which focus on other important features of life~\cite{ML21,smith2000origins,sole2024fundamental}. For instance, it differs from perspectives that emphasize the importance of Darwinian evolution and genetic information in living organisms~\cite{godfrey-smithBiologicalInformation2007,CA24,smith2000concept,mayrWhatEvolution2001}. In particular, unlike genetic information, which emerges at the level of replicator populations, semantic information operates in individual organisms, proto-organisms, or other self-maintaining systems. Our perspective is also different from those that focus on thermodynamic aspects of life~\cite{Sch44,morowitz2007energy}, such as metabolic cycles and maintenance of nonequilibrium states. In principle, some non-living systems (e.g., hurricanes) may be capable of maintaining stable nonequilibrium states without having semantic information about their environments. 

\begin{figure*}[t!]
    \centering
    \includegraphics[width=0.9\textwidth]{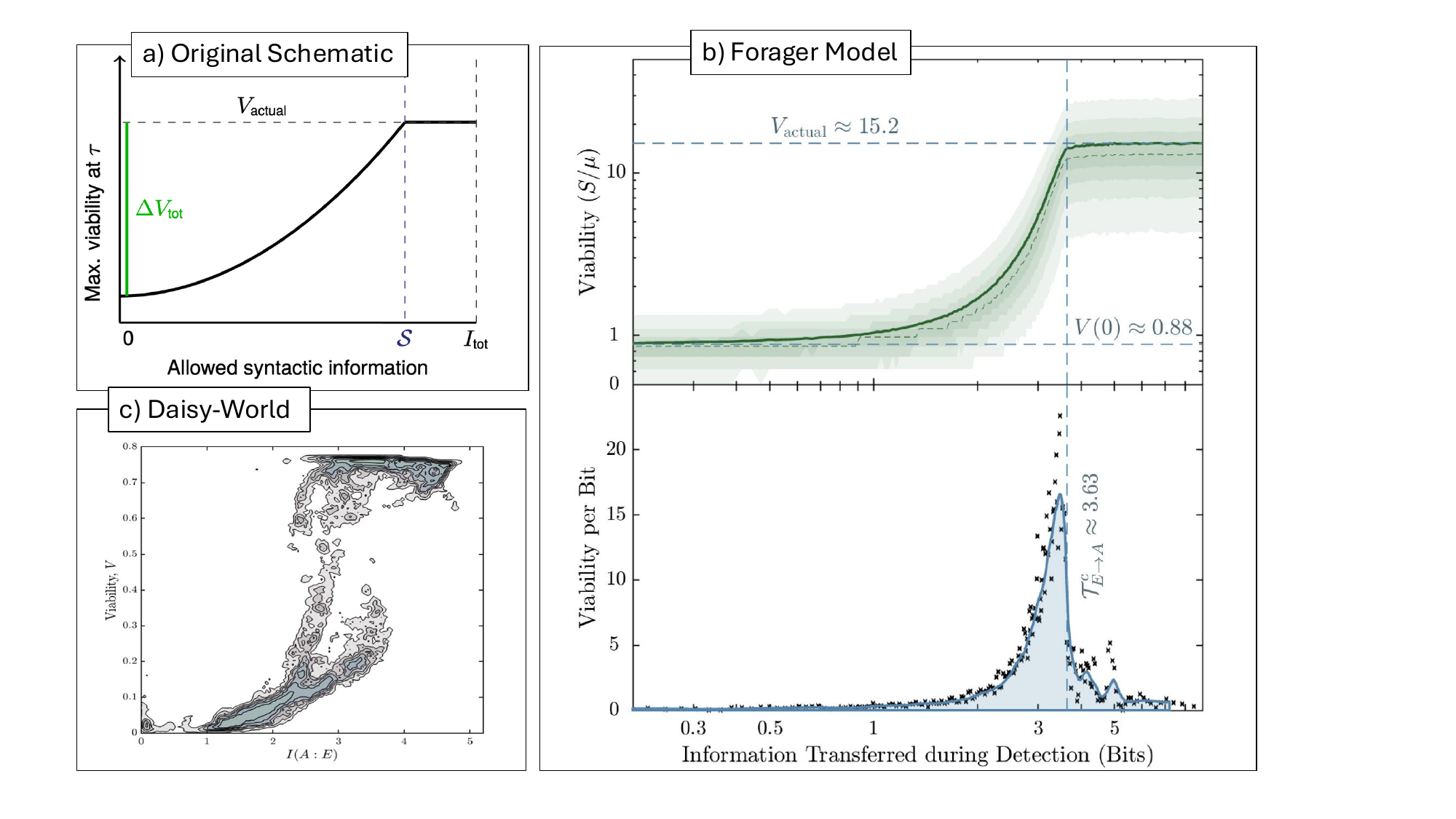}
   \caption{{\bf Semantic thresholds.} Semantic Information identifies ``semantic thresholds'' where only specific mutual information between an agent and its environment impacts viability. ({\bf a}) The foundational study~\cite{kolchinsky2018semantic} introduced this concept through counterfactual simulations assessing viability as information is scrambled. ({\bf b}) Forager models~\cite{Sowinski:2023vf} demonstrated such thresholds by adding noise to food detection sensors, revealing a peak in viability per bit of information at the threshold. ({\bf c}) Similar thresholds appear in biosphere models like Daisy-World, responding to stellar forcing~\cite{Sowinski2024eDW}.
   }    \label{fig:basics}
\end{figure*}

In the following, we discuss some emerging information-theoretic formalisms for investigating the informational perspective. We also discuss implications for fields such as the origin(s) of life and astrobiology, and promising directions for future work. This paper builds on ideas explored in a workshop on \emph{Information-Driven States of Matter}, held at the University of Rochester in July 2024. Bringing together more than 20 researchers, the workshop facilitated interdisciplinary discussions on how the informational perspective can unify investigations into principles of living matter, spanning fields like artificial life, biophysics, origin of life, and astrobiology.

\section{Theoretical frameworks}

\subsection{Semantic information}


The informational perspective calls for a mathematically precise and scientifically applicable theory of \textit{semantic information}. One such theory was introduced in Ref.~\cite{kolchinsky2018semantic}. The approach begins by defining state spaces and probability distributions for an agent $X$ and its environment $Y$, with correlations  quantified by the mutual information $I(X;Y)$. The agent's ability to maintain its own existence is quantified by a ``viability function''. 
By scrambling portions of the mutual information between the agent and its environment, the viability function quantifies which correlations are essential for survival. Semantic information refers to that part of the overall mutual information that influences the agent's viability. 

\begin{figure*}[t!]
    \centering
    \includegraphics[width=.8\textwidth]{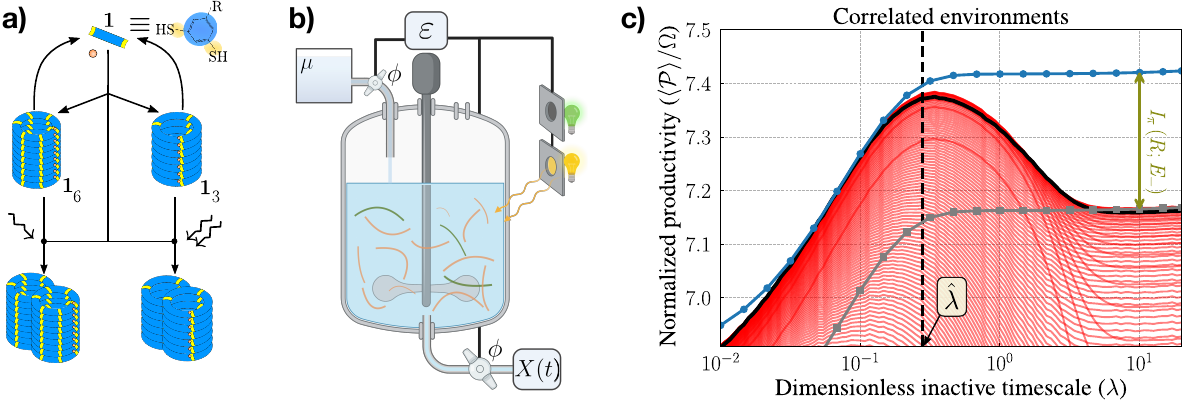}
   \caption{{\bf Semantic information in a flow reactor.} A proposed experiment for studying semantic information inspired by the ``fitness value of information'' from Ref.~\cite{pinero2024information}. (\textbf{a}) The system consists of a population of simple replicators, such as the photocatalytic molecular replicators synthesized by Liu et al.~\cite{liu2024light}. (\textbf{b}) The replicators are placed in a flow reactor and subjected to fluctuating environments which favor different replicators, e.g., weak (green) light or strong (orange) light. The population is also allowed to re-equilibrate during ``inactive'' periods (no growth). 
   (\textbf{c}) Productivity (replicator production per time) depends on environmental statistics as well as internal parameters, such as exchange reactions and re-equilibration timescale $\lambda$. At intermediate timescales, the system's memory may provide a source of side information, increasing productivity in proportion to the mutual information between successive environments~\cite{pinero2024information}.
   }    \label{fig:flowreactor}
\end{figure*}
Importantly, the viability function is not an externally imposed utility function, but rather it is defined an emergent property of the intrinsic dynamics of an agent coupled to an environment~\cite{kolchinsky2018semantic,rovelli_meaning_2018}. For instance, viability may be defined in terms of the agent's ability to resist equilibration (maintain low entropy) or to avoid absorption into a death-like attractor.

Recent studies have operationalized this framework in practical systems, see Figure~\ref{fig:basics}. For example, in foraging models~\cite{Sowinski:2023vf}, viability was quantified as the expected lifetime of a forager. By introducing noise into the forager's sensory inputs, semantic information was quantified as the subset of environmental correlations critical for maintaining viability. One of the key insights from~\cite{Sowinski:2023vf} is the discovery of a \textit{viability threshold}---a plateau in the viability curve where certain correlations between an agent and its environment do not influence survival. Below this threshold, information remains purely syntactic, devoid of semantic value. Above the threshold, survival declines monotonically as noise increases, indicating that only a subset of environmental correlations holds semantic significance.  This result underscores the critical distinction between syntactic and semantic information. 
Similar thresholds have also been identified in other contexts. In the Daisy-World model~\cite{Sowinski2024eDW}, such thresholds emerged within the biosphere-planet feedback mechanisms that regulate planetary conditions, relating semantic information to ecological stability. In networks of coupled Kuramoto-style oscillators~\cite{sowinski2024information}, the emergence of semantic thresholds was shown to depend on the underlying network topology, emphasizing how both structural and dynamical aspects shape semantic information.

Recent work has proposed extending these theoretical insights into experimental domains. Synthetic cells have been proposed as a novel platform to investigate semantic information~\cite{Magarini_2021}. Synthetic biology and molecular communication techniques enable the construction of programmable systems that encode specific chemical signals capable of inducing functional changes in a receiver system~\cite{rhee2012application,NEH,FVE16}; such systems provide a way to study semantic information (goal-oriented changes triggered by a message) by observing controlled self-organization and adaptive responses in synthetic cells. Active matter provides another promising platform for experimental investigation of semantic information theory, given recent work on minimal non-biological particles that can acquire and process information~\cite{hongChemotaxisNonbiologicalColloidal2007,liebchenSyntheticChemotaxisCollective2018,bauerleSelforganizationActiveParticles2018,ziepkeMultiscaleOrganizationCommunicating2022}.


\subsection{Fitness value of information}

As mentioned, organisms gather and process information to adapt their features and behaviors, enhancing viability and reproduction in dynamic conditions. 
One way in which this can occur is when populations deploy different phenotypes in order to ensure survival in fluctuating environments~\cite{yoshimura1996evolution}.

When the fit between phenotypes and environments affects multiplicative growth rate, the optimal population strategy in uncertain environments is {bet-hedging}, which was first derived information-theoretically by Kelly~\cite{kelly1956new}. Furthermore, when the phenotypic response can depend on an external signal or cue $Z$~\cite{donaldson2010fitness,rivoire2011value}, the mutual information $I(Z; Y)$ between environment states $Y$ and cues $Z$
controls the increase of maximal growth rate. In such cases --- which sometimes go under the name of \emph{fitness value of information} in the literature~\cite{donaldson2010fitness,rivoire2011value} --- there is a direct quantitative relationship between information and functional outcomes  (growth)~\cite{moffet2020fitness,marzen2018optimized,pinero2024information}.  In this sense, the fitness value of information can be seen as a special kind of semantic information.


This approach is further generalized by \textit{rate-distortion theory}~\cite{berger1971}, which introduces a {distortion function} $d(x, z)$ to quantify the cost of mapping environment  $x$ to cue $z$.  For example, $d(x, z)$ might represent the negative log growth rate when an organism observes cue $z$ in environment $x$. 
While bet-hedging directly maximizes the growth rate~\cite{donaldson2010fitness}, rate-distortion theory provides a broader framework for analyzing diverse strategies. For instance, an organism requiring multiple essential but non-interchangeable nutrients assigns different values to sensing each nutrient based on its current state~\cite{barker2024fitness,barker2022subjective}.  This framework has been considered both in information-theoretic and biophysical contexts~\cite{berger2002living,rivoire2011value,moffett2022minimal},  and it suggests that living systems optimize rate-distortion limits~\cite{moffett2021code,bialek2012biophysics}. Organisms near these limits exhibit efficient trade-offs, using  limited resources to harvest semantic information that most contributes to increasing growth~\cite{bialek2012biophysics}. 

Promising experimental tools include chemostats (continuous flow reactors), which measure growth rates in controlled environments~\cite{hoskisson2005continuous,wright2020single}. Such systems could be adapted to study the fitness value of information in synthetic and biological systems.  
In this setting, a recent preprint~\cite{pinero2024information} proposed a theoretical and experimental framework for studying the trade-offs between information and replicator production, see Figure~\ref{fig:flowreactor} for experimental setup. The framework is applicable to modern microbial organisms as well as minimal molecular replicators. 


\section{Implications for origins of life research}


The informational perspective 
shifts the focus of origin-of-life research. Rather than emphasizing the emergence of specific molecules or structures (e.g., RNA, ribosomes, or metabolic pathways), it emphasizes the transition from information-neutral systems to systems with semantic information --- that is, systems that maintain their viability by sensing and responding to their environments~\cite{barandiaran_normestablishing_2013,egbert2023behaviour}.  Identifying the necessary and sufficient conditions for such transitions suggests new possibilities for understanding abiogenesis, not only on Earth~\cite{LNB24} but also on Mars, subsurface ocean worlds, exoplanets, and even in artificial systems~\citep{bartlett2020defining,ML21}. 

In fact, models of emergent learning and adaptive behavior in protocells or chemical systems illustrate how simple systems may leverage information for functional purposes. 
In other words, even simple systems can exhibit minimal semantic information, responding to environmental perturbations in ways that prolong their existence~\cite{egbert2023behaviour}, thereby suggesting that semantic information may have emerged during the early stages of life, potentially facilitating other critical transitions in abiogenesis~\cite{jeancolas2020thresholds,ML21}. If extant life on Earth is one of many possible instantiations of living systems~\cite{LNB24}, then extraterrestrial life and even digital life could represent additional members of this category~\citep{bartlett2020defining,ML21}, and could evince similar features.



Recent proposals suggest that origins experiments focus on detecting complexity and information processing, rather than specific molecules or replication~\cite{bartlett2019probing}. This could involve testing whether a system adapts to changing conditions by increasing its ``statistical complexity''~\cite{crutchfield2012between}, indicating the ability to process and learn from information. Such learning has been observed in diverse systems~\cite{buckley2024natural,zhong2021machine}, and this information-driven approach could guide experimental and computational investigations into origins~\cite{duenas2019chemistry,BL22}.

\begin{figure}[t!]
    \centering
    \includegraphics[width=.9\columnwidth]{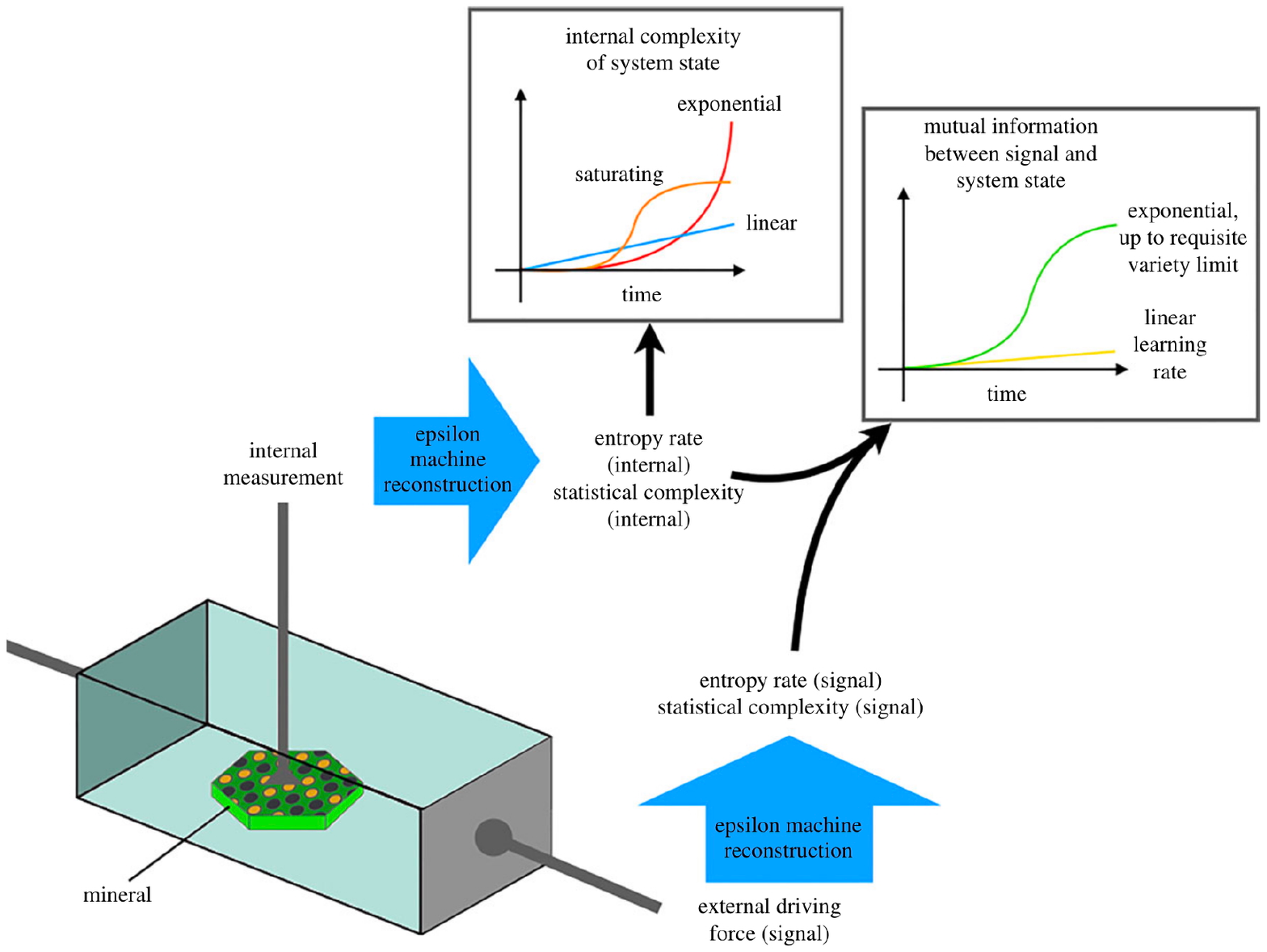}
    \caption{{\bf Example schematic for an origins of life experiment} focusing on information measures. Consult the text for full details. Reproduced with permission from~\citep{bartlett2019probing}.}
    \label{fig:info_origins}
\end{figure}

For example, Figure~\ref{fig:info_origins} presents a potential origins-of-life experiment, which focuses on information and complexity. In this setup, a growing chemical garden is driven away from equilibrium by electrodes, which produce a time-varying electrochemical gradient across the system; note that such gradients may have been crucial in facilitating the origin of life~\cite{BBG17,NSD23}. The variability of this electrochemical potential is controlled automatically according to an \emph{epsilon machine}. An epsilon machine is a hidden-variable dynamical model with associated measures of dynamical complexity, such as statistical complexity and entropy rate~\cite{crutchfield2012between}. Measurements of internal state variables are taken, producing a time series. This time series also has an associated statistical complexity and entropy rate, derived through the algorithm of ``epsilon machine reconstruction''.

The premise of this experiment is to see whether the system can transition to being information-driven, if we incrementally increase the complexity of the external driving force, and observe whether the internal complexity changes as well. If no such information transition happens, there will be little to no mutual information between the system's response and the driving force (the electrochemical gradient). On the other hand, if information, or perhaps semantic information, becomes a determining factor in this system, then the internal statistical complexity will track the external statistical complexity in some way (the system will learn)~\citep{zhong2021machine}. This learning may be approximately linear (internal statistical complexity is a linear function of time, if the external statistical complexity increases linearly with time), or it might exhibit more exotic dynamics, such as starting out slow and then exhibiting exponential behavior, before saturating into an overall sigmoid. This could be due to the system learning as much as is feasible, reaching the so-called limit of requisite variety~\citep{klir1991requisite}.

In this experiment, information may be said to drive a system's dynamics once certain conditions are met: (a) the environment presents learnable features, (b) the system iscapable of information processing (e.g., protocells with basic learning capabilities~\cite{BL22}), and (c) feedback mechanisms connect information processing to increased viability. These mechanisms could enhance access to free energy~\cite{pinero2024optimization} or other resources for maintaining a nonequilibrium state, such as recognizing temporal patterns in energy availability or avoiding adverse conditions. 


\section{Implications for astrobiology}

Astrobiology is a rapidly emerging field dedicated to addressing the profound question: \textit{are we alone?}~\cite{SMI18,Co20,MA24}. Defined by NASA as the study of the ``{origin, evolution, distribution, and future of life in the Universe}'' \footnote{\url{https://astrobiology.nasa.gov/research/astrobiology-at-nasa/exobiology/}}, the field aims to uncover life’s potential elsewhere. Two main areas of research include understanding planetary \textit{habitability} --- the potential of different  extraterrestrial environments to support life~\cite{LBC09,CBB16,SMI18,SK21,ML21} --- and developing reliable methods for identifying \textit{biosignatures}, the indicators of extant or extinct life~\cite{NHV18,CW19,ML21,SL24}.

\begin{figure}[t]
    \centering
    \includegraphics[width=.9\columnwidth]{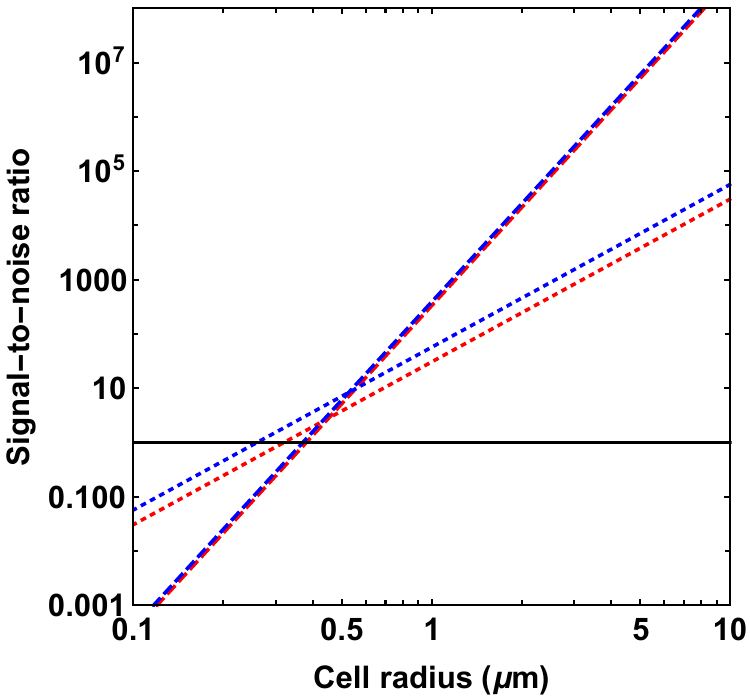}
   \caption{{\bf Signal-to-noise (SNR) associated with chemosensing} plotted as a function of cell radius for Earth's oceans (blue) and the hydrocarbon lakes of Saturn's moon, Titan (red); the dotted and dashed lines correspond to spatial and temporal modes of sensing~\citep{LM21}. A heuristic minimum cell radius may be inferred by determining when the SNR exceeds the value of unity (horizontal black line).
}    \label{fig:habitability_thresholds}
\end{figure}

In the astrobiological context, the informational perspective elucidates the constraints and possibilities for the existence and persistence of life, thereby adumbrating a unifying framework that may apply both on Earth and in extraterrestrial settings. This section explores how these informational constraints shape planetary habitability and guide the quest for biosignatures, contributing to a broader theory of life's distribution in the universe.

\subsection{Informational Constraints on Habitability}\label{SSecInfHab}

As discussed above, information processing is a universal feature of living systems, extending plausibly to extraterrestrial life. Two central facets of this capability are the abilities to sense~\cite{AGZ,HT14,BS18} and transmit~\cite{RCR50,WRL99,FVE16} information. Microorganisms, with their simplicity and ubiquity, provide a key case study for understanding these processes.

Mechanisms like chemotaxis, phototaxis, and thermotaxis allow microbes to detect gradients of chemicals, light, and temperature, respectively, offering critical insights into their environments~\cite{Ar99,WA04,WCV18}. These sensory capabilities enable behaviors such as nutrient acquisition, toxin avoidance, biofilm formation, and symbiosis~\cite{PWA11,Kir18,WCV18,RFL19}. In parallel, molecular communication through signaling molecules~\cite{DL90,RCL12,NEH} fosters cooperation, cognition, and multicellularity~\cite{WRL99,BLB02,WB05,EBJ09,PL15,GW20,JAS21}. These processes underscore the centrality of information in driving the evolution and adaptability of life.

Potential habitats for the origins, evolution, and sustenance of life span diverse physicochemical conditions, ranging from hydrothermal vents to soda lakes and pumice rafts~\cite{SAB13,SGS20,ML21,WBF23,RAH24} on Earth. Understanding how environmental parameters (e.g., temperature, viscosity) shape informational constraints is crucial for discerning which settings are conducive to the existence and sustenance of information-driven life. 

This intersection of biophysics and astrobiology remains distinctly underexplored~\cite{ML21}, with explicit connections to planetary environments just beginning to emerge~\cite{scharf2024rebuilding,MA24}. For instance, Ref.~\cite{LM21} modeled the lower bounds on cell sizes capable of gradient sensing and motility in various environments. Using signal-to-noise ratio (SNR) calculations, they predicted a theoretical chemosensing limit of $R_\mathrm{min} \sim 0.3$–$0.4$ $\mu$m in Earth’s oceans, consistent with empirical bounds of $\sim 0.4$ $\mu$m~\cite{MWJ15}. Similar calculations for Titan’s hydrocarbon lakes revealed comparable thresholds, despite the extreme conditions. Figure~\ref{fig:habitability_thresholds} illustrates these size limits and their implications for different planetary conditions.

Another study~\cite{ML24} estimated data rates ($\mathcal{I}$) for molecular communication via signaling molecules between microbes across habitats. $\mathcal{I}$ and its dependence on intercellular distances ($d$) were demonstrated to be broadly consistent with certain laboratory experiments \cite{GKP18} and numerical simulations \cite{SAC23}.  The findings of this work --- spanning Earth’s upper oceans ($\mathcal{I} \sim 4.7 \times 10^{-2}$ bits/s) to Titan’s lakes ($\mathcal{I} \sim 2.6 \times 10^{-1}$ bits/s) --- illustrated how $d$ and environmental variables influence informational dynamics, and were subsequently harnessed to explore ramifications for Earth's biosphere~\cite{LFB23}. 

This kind of studies highlight the putative major role of informational constraints in shaping habitability.

\subsection{Information-Centric Biosignatures}\label{SSecInfBioSig}

Agnostic biosignatures, which are designed in principle to transcend Earth-centric assumptions, have emerged as a promising tool in astrobiology~\cite{CN01,CHP19,MA24}. These approaches prioritize universal features of life over specific molecular markers, addressing the challenges of abiotic false positives~\cite{CW19,BRW22,MTB23}. For instance, abiotic processes can mimic cell-like morphologies~\cite{GRC02,MC22}, complicating microfossil identification. Agnostic biosignatures aim to avoid such pitfalls by focusing on life’s core characteristics, making them particularly valuable for detecting non-Earth-like life forms~\cite{bartlett2020defining,MA24}.

Proposals for agnostic biosignatures encompass both \emph{in situ} analyses and remote sensing. These include binding pattern analysis of nucleic acid molecules~\cite{JAG18}, assembly theory for molecular construction steps~\cite{MMC21}, and machine learning for distinguishing abiotic and biological materials~\cite{CHP23}. Information-centric approaches, such as epsilon machine reconstruction~\cite{BLG22} and Jensen-Shannon divergence for spectral analysis~\cite{VGK24} have likewise shown promise in identifying biosignatures through time series data and atmospheric spectra.

\emph{In situ} strategies, such as the ``poke it and look for a response'' method, could leverage dynamic responses to stimuli to differentiate living systems from abiotic materials. Possible measures in this context include information flow \citep{TS00,HPV07} and information efficiency \citep{AJM09,BHS14}, each of which can perhaps distinguish living organisms from dead ones or abiotic materials. 
By integrating such approaches with informational constraints on habitability, astrobiologists can refine their methods to account for the unique challenges posed by extraterrestrial environments. For example, as discussed earlier, cell size thresholds derived from information sensing models can serve as heuristic filters in biosignature evaluation.


\section{Summary and future directions}

This paper explores the informational perspective, which posits that life's unique ability to acquire, process, and utilize information as one of the important aspects that distinguish it from non-life. In this brief piece, we discussed formal approaches, such as semantic information and fitness value of information, for quantifying those aspects of information that contribute to agent viability.  We also discussed implications for ongoing research in fields like origins-of-life and astrobiology.

The informational perspective complements existing perspectives on essential features of life by focusing on underlying principles rather than its specific manifestations. It contributes to the broader dialogue on what defines life and how it persists and evolves, on its origins and its potential existence beyond Earth. 


Further development and integration of theoretical and experimental approaches is key to further progress.  Advances in active matter, synthetic biology, and molecular communication systems offer powerful tools for constructing and analyzing systems that mimic the informational dynamics of natural organisms, including processes like learning, adaptation, and feedback in controlled environments. Such experiments may not only test theoretical predictions but also uncover novel information-driven phenomena that may have played a role in the origin and evolution of life. 


\begin{acknowledgements} 
GG, AF and AK acknowledge support through Grant No. 62417 from the John Templeton Foundation. The opinions expressed in this publication are those of the author(s) and do not necessarily reflect the views of the Foundation. AK was partly supported by the European Union's Horizon 2020 research and innovation programme under the Marie  Sk{\l}odowska-Curie Grant Agreement No.~101068029. ME acknowledges and appreciates the support of the University of Auckland's Research and Study Leave Programme. AE acknowledges support from Discovery grant RGPIN-2016-05288 from the Natural Sciences and Engineering Research Council (NSERC). 
\end{acknowledgements}


%

\end{document}